\documentclass[aps,prb,twocolumn,superscriptaddress,showpacs]{revtex4-1}

\usepackage{amsmath}
\usepackage{graphicx}
\usepackage{epstopdf}
\usepackage{epsfig}
\usepackage{amsfonts}
\usepackage[breaklinks=true]{hyperref}
\usepackage{hypcap}
\usepackage{verbatim}
\usepackage{multirow}
\usepackage{tabularx}
\usepackage{color}
\usepackage{natbib}
\usepackage{bigstrut}
\usepackage{array}

\newcolumntype{C}{>{\centering\arraybackslash}X}%
 
\newcolumntype{x}[1]{>{\hfil$\displaystyle} p{#1} <{$\hfil}} 
\def\beq{\begin{equation}}
\def\eeq{\end{equation}}

\newcommand{\bmat}{\left(\begin{matrix}}
\newcommand{\emat}{\end{matrix}\right)}

\newcommand{\tp}{\widetilde{\Psi}}

\newcommand{\sgn}{\operatorname{sgn}}
\newcommand{\eq}[1]{Eq.~(\ref{#1})} 
\newcommand{\fig}[1]{Fig.~\ref{#1}} 

\begin{document}

\title{Conductance spectroscopy of topological superconductor wire junctions}
\author{F. Setiawan}
\author{P. M. R. Brydon}
\author{Jay D. Sau}
\author{S. Das Sarma}
\affiliation{Condensed Matter Theory Center and Joint Quantum
  Institute, Department of Physics, University of Maryland, College
  Park, Maryland 20742-4111, USA} 
\date{\today}

\begin{abstract}
We study the zero-temperature transport properties of one-dimensional
normal metal-superconductor (NS) junctions with topological superconductors across their topological
transitions. Working within the Blonder-Tinkham-Klapwijk (BTK) formalism generalized for topological NS junctions, we
analytically calculate the differential conductance for tunneling into
two models of a topological superconductor: a spinless intrinsic $p$-wave
superconductor and a spin-orbit-coupled $s$-wave superconductor in a
Zeeman field. In both cases
we verify that the zero-bias conductance is robustly quantized at
$2e^2/h$ in the topological regime, while it takes nonuniversal
values in the nontopological phase. The conductance spectra in the topological state
develops a peak at zero bias for certain parameter regimes, with 
the peak width controlled by the strength of spin-orbit coupling and barrier transparency. 
\end{abstract}

\maketitle

\section{Introduction}

The search for topological superconductors in solid-state systems is
motivated by the possibility of realizing Majorana zero-energy modes
at their surfaces, which are of both fundamental and technological
importance~\cite{ReadGreen2000,kitaev2001,NayakRMP2008}. In the absence of intrinsic topological superconductors,
much effort has been put into engineering such systems from
conventional
components~\cite{sau10a,sau10b,lutchyn10,oreg10,alicea10}. In addition
to the proposals involving semiconductor-superconductor hybrid
structures~\cite{sau10a,sau10b,lutchyn10,oreg10,alicea10} as hosts for
Majorana modes, which have attracted considerable experimental
attention~\cite{mourik12,Rokhinson12,deng12,das12,Finck2013,churchill13,Lee13},
there have been many recent theoretical proposals for artificially
engineering effectively spinless low-dimensional $p$-wave topological
superconductors~\cite{Fu08,Fu09,Mi13,Choy11,Zhang08,Sato09,Duckheim11,Chung11,Takei11,Mao12,Sau12,Kim14,Brydon15,Hoi15},
which could localize zero-energy Majorana modes at suitable defects
such as vortex cores or system boundaries. The subject has been
extensively reviewed in the recent
literature~\cite{Alicea12,Tudor13,DasSarma15,Franz15}.  

It is particularly desirable to realize spinless $p$-wave
superconductors, as they support a single Majorana mode at their
boundaries~\cite{ReadGreen2000,kitaev2001}. The most promising of
these proposals involves proximity-inducing superconductivity in a 
spin-orbit-coupled semiconducting nanowire in the presence of a
magnetic field~\cite{lutchyn10,oreg10,alicea10}, which has
subsequently been the
subject of a number of experiments~\cite{mourik12,Rokhinson12,deng12,das12,Finck2013,churchill13,Lee13}. By
varying the magnetic field, the system is predicted to undergo a
transition from a nontopological to  a topological phase. Such an external magnetic-field-induced topological quantum phase transition has the considerable advantage of tuning the existence (or absence) of the Majorana mode in the experimental system simply by changing the Zeeman field. A key
signature of the topologically nontrivial state is the
quantized value $2e^2/h$ of the differential conductance for tunneling
into the wire at zero-bias voltage. This
quantized conductance, associated with perfect Andreev reflection, indicates the presence of a single localized
Majorana zero-energy mode at the wire
end~\cite{Sengupta01,Flensberg10,Law09,Wimmer11,Xin15}. For a sufficiently  
high tunnel barrier, the conductance spectra will be
peaked with this value at zero bias. While experimental results
clearly show the development of such a peak upon tuning the system, at a finite magnetic field, into the predicted 
topological regime, the 
value of the zero-bias conductance peak is much less than the expected quantized value. The
reasons for this discrepancy are addressed in Refs.~\cite{Lin12,Pientka12}, and alternative nontopological explanations have been
advanced~\cite{Kells12,Liu12,Bagrets12,Pikulin12,Rainis13,Roy13}. The lack of quantization of the experimental observations can be reconciled~\cite{sau10b,Lin12} with the Majorana theory by including the finite temperature and the finite length of the nanowire (thus allowing the Majorana modes from the two ends to overlap), but this physics is beyond the scope of our work where we restrict to zero temperature and a single normal metal-superconductor (NS) junction (assuming the other Majorana mode to be far away from this junction).  

The difficulty in interpreting the tunneling experiments has prompted
numerous theoretical studies on the conductance of the nanowire
device, using both numerical~\cite{Stanescu11,Prada12,Stanescu14,Roy13,Lin12,Qu11,Dibyendu12,Rainis13,Dibyendu13} and
analytical 
techniques~\cite{James14,Yan14a,Rex14,Thakurathi12,Yan14b,Sun15}.   
Although the latter works consider highly idealized models of the system,
they are nevertheless valuable as they give clear insight into the
parametric dependence of the transport physics as well as its dependence on various physical properties of the experimental setup, which can then be
applied to understand the more  
complicated numerical studies. An important question
concerns the change in the conductance as the system is tuned from the
topologically trivial to the nontrivial regimes (e.g., by tuning the applied magnetic field in semiconductor-superconductor hybrid structures). Remarkably, this
aspect of the physics has attracted relatively little attention using
these analytic methods~\cite{Rex14}. The purpose of this paper is to
analytically address this aspect of Majorana physics in topological nanowire junctions.

In this paper we examine the conductance spectra of one-dimensional
NS junctions involving topological superconductors across their
topological transition. We utilize the
Blonder-Tinkham-Klapwijk (BTK) formalism~\cite{BTK}, 
  which is commonly employed to study junctions with unconventional
  superconductors~\cite{Tanaka94,Tanaka00,tanaka10,takami}, to obtain analytic
results for the tunneling conductance of two models of a topological
superconductor junction: a junction between a spinless normal metal
and a $p$-wave superconductor, and a junction between a
spinful normal metal and a spin-orbit-coupled $s$-wave
superconductor in a magnetic field. The former is the simplest model
for tunneling into a topological
superconductor~\cite{Yan14a,Thakurathi12,Yan14b}, while the latter is
a minimum model~\cite{sau10b} for the semiconductor nanowire device
where experimental signatures for Majorana zero modes have been
reported through the observation of zero-bias tunneling conductance
peaks at the NS junction. We note that the 
spinless $p$-wave superconductor can be regarded as an effective
low-energy theory for the semiconductor nanowire, but this is
inadequate for understanding the conductance spectroscopy of the
device. Our analysis is analytical, and in particular we give explicit
expressions for the zero-bias tunneling conductance at zero
temperature, which clearly shows an abrupt change at the topological
transition. Specifically, we find that in the topological regime, the
zero-temperature zero-bias 
conductance is quantized at a value of $2e^2/h$ independent of the
barrier strength $Z$, but the detailed structure (e.g., the width and the shape) of the quantized zero-bias
  conductance peak is controlled by the barrier transparency and the
  magnitude of spin-orbit coupling. Our BTK theory for the topological
  NS junction also shows that a finite barrier transparency could lead
  to the experimentally observed soft gap which is ubiquitous in
  semiconductor nanowire tunneling
  experiments~\cite{mourik12,das12,Finck2013,churchill13}. 

The paper is organized as follows. In Sec.~\ref{sec:np}, we warm up by
studying the
conductance of a junction between a spinless normal metal and a
spinless $p$-wave superconductor across the topological 
transition. We then generalize the theory to consider the
semiconductor nanowire device in Sec.~\ref{sec:ns}. In particular, we
obtain analytic results for the conductance spectra in the limits of
a strong Zeeman field and strong spin-orbit coupling. Finally, the
results are summarized in Sec.~\ref{sec:summary} with a conclusion. 

\section{Junction with a spinless $p$-wave superconductor}\label{sec:np} 

We start by considering a one-dimensional junction between a spinless
normal metal (NM) and a $p$-wave superconductor ($p$SC), which are
located at $x\leq0$ and $x\geq0$, respectively. Their interface at
$x=0$ is modeled by a $\delta$-function barrier of strength $Z$
following the BTK prescription.  The parameter $Z$ controls the
barrier transparency at the NS interface, and is the key parameter in the
theory quantifying the tunneling conductance properties at the
junction: a low (high) value of $Z$ corresponds to a barrier with high
(low) 
transparency at the NS interface. A microscopic evaluation of $Z$
is typically difficult since the microscopic details of the junction
are generally unknown, and so $Z$ is treated as a free fitting parameter. The 
Hamiltonian in each region is written $ H_j(x)= \frac{1}{2}\int 
dx\Psi^{\dagger}_j(x)\mathcal{H}_j(x)\Psi_j(x)$, where $\Psi_j(x) = 
(\psi^\dagger_j(x),\psi_j(x))^{\mathrm{T}}$ are Nambu spinors and 
$\psi^\dagger_j(x)$ $(\psi_j(x))$ denotes the creation (annihilation)
field operator in region $j= N$ (NM) and $p$ ($p$SC). Assuming that the mass
$m$ is uniform throughout the system, the
Bogoliubov-de Gennes (BdG) Hamiltonians are 
\begin{subequations}\label{eq:hnp}
\begin{align}
\mathcal{H}_N(x) &= (-\hbar^2\partial_x^2/2m- \mu_N)\tau_z, \\
\mathcal{H}_p(x) &=(-\hbar^2\partial_x^2/2m- \mu_p)\tau_z -i\Delta_p \partial_x\tau_x,
\end{align}
\end{subequations}
where $\mu_N$ ($\mu_p$) is the chemical potential of the NM ($p$SC),
$\Delta_p \geq 0$ is the $p$-wave pairing potential, and
$\tau_{\mu}$ are the Pauli matrices acting on the particle-hole
space.

\begin{figure}[h]
\capstart
\begin{center}
\includegraphics[scale=0.7]{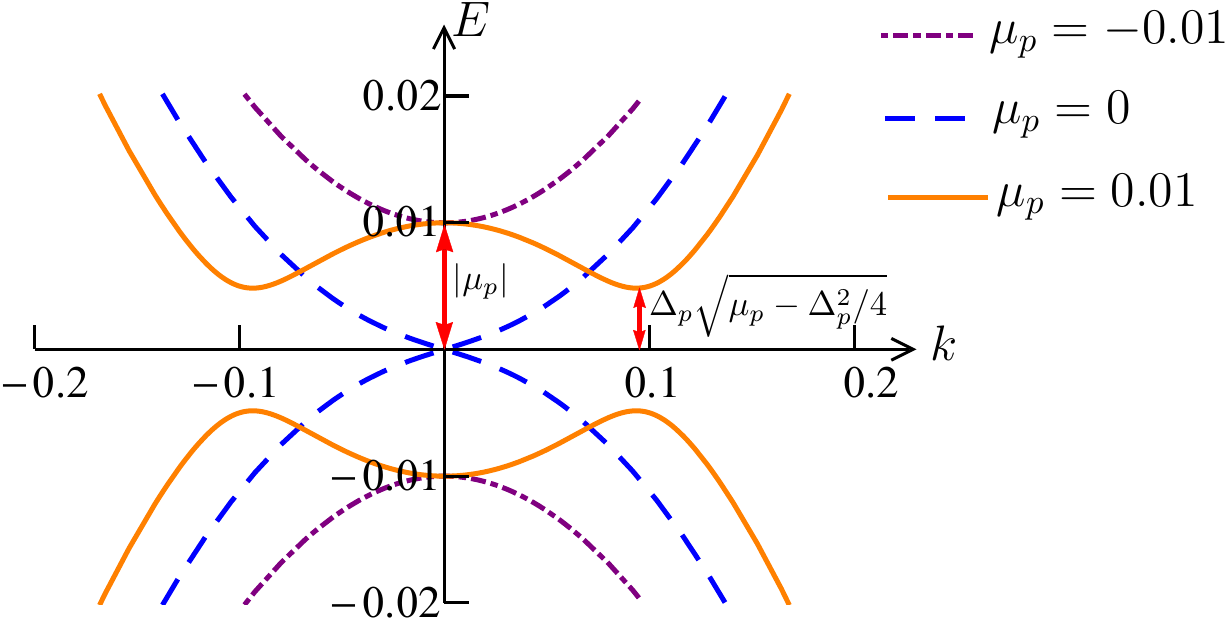}
\end{center}
\caption{(Color online) Typical energy spectra of the spinless $p$SC
  illustrating the nontopological 
  ($\mu_p = -0.01$), transition ($\mu_p = 0$) and topological regimes
  ($\mu_p = 0.01$). In all curves we set $\Delta_p = 0.05$.}\label{fig:ep} 
\end{figure}

For notational simplicity, in the following we work with units such
that $\hbar$, $\mu_N$, and $2m$ are all equal to unity. The energy
spectra of the NM and $p$SC  
are then given by
$\epsilon_{N,\pm}(k) = \pm(k^2 -1)$ and $\epsilon_{p,\pm}(k) = \pm\sqrt{(k^2 - \mu_p)^2 + (\Delta_p
  k)^2}$, respectively. In~\fig{fig:ep} we plot the spectrum of the
$p$SC for different values of $\mu_p$. Note that the spectrum becomes
gapless at $\mu_p = 0$ which marks the topological transition~\cite{ReadGreen2000} between
the BEC-like strong pairing phase ($\mu_p <0$) and the
BCS-like weak pairing  phase ($\mu_p >0$). In the latter case, the
positive energy spectrum only develops the characteristic ``double-well''
BCS structure for  $\mu_p > \Delta_p^2/2$, with 
minimum value $E_{1} = \Delta_p\sqrt{\mu_p-\Delta_p^2/4}$ at $k =
\pm\sqrt{\mu_p-\Delta_p^2/2}$, and a local maximum value $E_{2} =
\mu_p$ at $k=0$.  

We consider the scattering of an electron injected from the NM into
the $p$SC with energy $E$. The incident electron can be normal
reflected as an electron, Andreev reflected as a hole, or transmitted
into the $p$SC. The scattering wave function is $\Phi(x) =
\Phi_N(x)\Theta(-x) + \Phi_p(x)\Theta(x)$, where $\Theta(x)$ is the
Heaviside step function and  
\begin{subequations}
\begin{align}
\Phi_N(x)  &= \bmat 1 \\ a \emat e^{ix} +\bmat b\\ 0 \emat  e^{-ix} \,, \\
\Phi_p(x) &= c\bmat \gamma_-\\1 \emat e^{ik_-x} + d\bmat \gamma_+\\1 \emat e^{ik_+x},
\end{align}
\end{subequations}
where $a$ and $b$ are the Andreev and normal reflection amplitudes,
respectively, $c$ and $d$ are the transmission coefficients into
the $p$SC, and 
\begin{equation}\label{eq:component}
\gamma_{\pm} = \frac{E + k_{\pm}^2 - \mu_p}{\Delta_p k_{\pm}}.
\end{equation}
Note that
we approximate the wave vector of the electrons and holes in the NM by the Fermi momentum $k_F = \sqrt{2m\mu_N}/\hbar$, valid for $E\ll 1$. The momenta $k_{\pm}$
of the $p$SC wave function are solutions of the equation 
\begin{equation}\label{eq:ep}
 E^2 = (k^2 - \mu_p)^2 + (\Delta_p k)^2\,.
\end{equation}
Depending on the energy $E$ of the incoming electron and the chemical
potential $\mu_p$, the $p$SC wave function can either be
evanescent with complex solutions of~\eq{eq:ep}, or involve
propagating states corresponding to real solutions of~\eq{eq:ep} with
positive group velocity. We classify the different solutions in
Table \ref{tab:ksol}.  

The wave functions satisfy the continuity equation $\Phi_p(x)|_{x=0^+}
= \Phi_{N}(x)|_{x=0^-}$ and current conservation condition
$J_p\Phi_p(x)|_{x=0^+} - J_N\Phi_{N}(x)|_{x=0^-} = -2iZ\tau_z\Phi_N(0)$ 
where the current operators are given by 
\beq
J_N = -2i\partial_x\tau_z\,, \qquad
J_p = 
-2i\partial_x\tau_z +\Delta_p\tau_x\,.
\eeq
Solving the boundary conditions, we derive the Andreev and normal
reflection  coefficients 
\begin{widetext}
\begin{subequations}
\begin{align}
a(E) &= \frac{\Delta_p(\gamma_+ - \gamma_-)- 2(k_+ -
  k_-)}{\mathcal{D}_E}, \hspace{6 cm} \mathrm{and} \\
b(E) &= \frac{(2-2iZ-k_+ -k_--\Omega)(\gamma_- - \gamma_+) +\frac{\Delta_p}{2}(k_+ - k_-)(\gamma_+\gamma_- +
  1)}{\mathcal{D}_E}, 
\end{align}
\end{subequations}
respectively, where
\begin{subequations}
\begin{align}
 \Omega &= 1+ (Z - ik_-)(Z-ik_+)  - \frac{\Delta_p^2}{4}, \hspace{6 cm} \mathrm{and} \\
\mathcal{D}_E &= \Omega(\gamma_- - \gamma_+) -\frac{\Delta_p}{2}(k_+ -
k_-)(\gamma_+\gamma_- + 1) - (k_+ - k_-)(\gamma_- + \gamma_+). 
\end{align}
\end{subequations}
\end{widetext}

\begin{table}[h!]
\capstart
\begin{tabularx}{\linewidth}{|C|C|C|}
\hline
$\mu_p$ & $E$ & $k_-,k_+$ \\\hline
\multirow{2}{*}{$\mu_p\leq\Delta_p^2/4$} & $0 \leq E \leq E_2$ &
$k_{I-}$, $k_{I+}$ \\
& $E \geq E_2$ & $k_{I+}$, $k_{R+}$  \\ \hline
\multirow{3}{*}{$\Delta_p^2/4\leq\mu_p\leq\Delta_p^2/2$} & $0\leq E\leq E_1$ & $k_{C-}$, $k_{C+}$ \\
& $E_1\leq E \leq E_2$ & $k_{I-}$, $k_{I+}$ \\
& $E \geq E_2$ & $k_{I+}$, $k_{R+}$ \\\hline
\multirow{3}{*}{$\mu_p\geq\Delta_p^2/2$} & $0\leq E\leq E_1$ & $k_{C-}$, $k_{C+}$ \\
& $E_1\leq E \leq E_2$ & $k_{R-}$, $k_{R+}$ \\
& $E \geq E_2$ & $k_{I+}$, $k_{R+}$\\\hline
\end{tabularx}
\caption{Various solutions of Eq.~(\ref{eq:ep}) for different
  values of chemical potential $\mu_p$ and energy $E$, where $E_1
  = \Delta_p\sqrt{\mu_p - \Delta_p^2/4}$ and $E_2 =
  |\mu_p|$. We
  denote propagating solutions by $k_{R\pm}$, while
  evanescent solutions are given by $k_{I\pm}$ and
  $k_{C\pm}$. These are given by
  $k_{R\pm} = \pm[(\mu_p-\Delta_p^2/2) \pm \sqrt{E^2 - E_1^2}]^{1/2}$,
  $k_{I\pm} = i[(\Delta_p^2/2-\mu_p) \pm \sqrt{E^2 - E_1^2}]^{1/2}$
  and $k_{C\pm} = \pm[(\mu_p - \Delta_p^2/2) \pm i
    \sqrt{E_1^2-E^2}]^{1/2}$. }\label{tab:ksol} 
\end{table}

\begin{table}[h]
\capstart
\setlength{\tabcolsep}{4pt}
\renewcommand{\arraystretch}{3}
\begin{tabular}{|c|c|c|c|}
\hline
& $\mu_p < 0$ & $\mu_p = 0$ & $\mu_p > 0$ \\\hline
$a(0)$ & $0$ & $\displaystyle-\frac{i\Delta_p}{(Z + \Delta_p/2)^2 + 1 + \Delta_p}$ & $-i$ \\\hline
$b(0)$ & $-e^{i\varphi}$ & $\displaystyle-\frac{(Z + \Delta_p/2)^2 + 1}{(Z + \Delta_p/2)^2 + 1 + \Delta_p} e^{i\varphi}$ &0  \\\hline
$\displaystyle{\frac{G_p(0)}{G_0}}$ & $0$ & $\displaystyle 1-\frac{[(Z+\Delta_p/2)^2+1]^2-\Delta_p^2}{[(Z + \Delta_p/2)^2 + 1 + \Delta_p]^2}$ & 2\\\hline
\end{tabular}
\caption{Explicit expressions for the zero-bias Andreev reflection
  coefficient $a(0)$, normal 
  reflection coefficient $b(0)$, and differential conductance
  $G_p(0)$ for the spinless NM-$p$SC junction. The results are
  classified according to the three different regimes of $\mu_p$: the
  nontopological state ($\mu_p<0$), the topological phase transition
  point ($\mu_p=0$), and the topological state ($\mu_p>0$). The quantity $\varphi$
  is defined by $\sin\varphi =
  2(Z+\sqrt{\Delta_p^2/4-\mu_p})/[(Z+\sqrt{\Delta_p^2/4-\mu_p})^2+1]$.}\label{tab:np} 
\end{table}

Within the BTK formalism~\cite{BTK} the zero-temperature differential
conductance is given by  
\begin{equation}
G_p(E) = G_0\left(1 +|a(E)|^2 - |b(E)|^2\right),
\end{equation}
where $G_0 = e^2/h$ is the normal state conductance for a
quantum point contact. Although the general form of $G_p(E)$ is
lengthy and unenlightening, relatively simple expressions can be found
for the physically interesting case of zero bias, i.e., $E=0$, which we
provide in Table \ref{tab:np} for the three different regimes of
$\mu_p$. In particular, we find that the zero-bias conductance
abruptly jumps from 
$G_p(0) = 0$ in the trivial regime ($\mu_p<0$) to $G_p(0) = 2$ in the
topological regime ($\mu_p>0$). The quantized conductance is
characteristic of the topological state, and can be interpreted as
indicating perfect Andreev reflection [i.e., $|a(0)|^2 =1$, 
$|b(0)|^2=0$] at an interface supporting a Majorana
mode~\cite{Law09,Flensberg10,Xin15}. It is 
therefore independent of the barrier strength $Z$ and
$p$-wave pairing potential $\Delta_p$. At
the transition point ($\mu_p=0$) we find $G_p(0) \leq G_0$, with
the exact value depending upon $Z$ and $\Delta_p$.

\begin{figure}[h!]
\capstart
\begin{center}
\includegraphics[width=\linewidth]{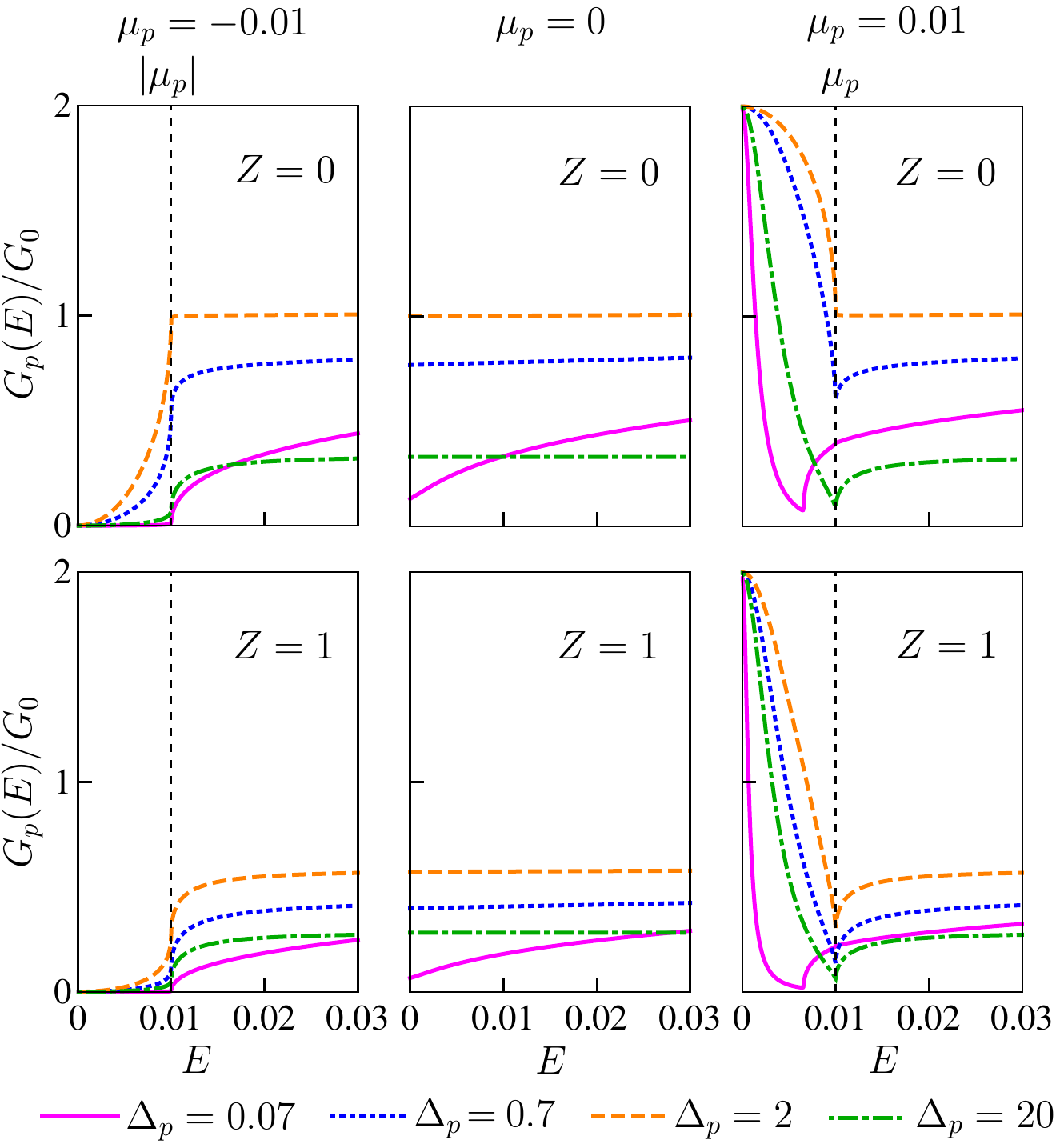}
\end{center}
\caption{(Color online) Variation of the tunneling conductance 
  $G_p(E)$ with pairing potential $\Delta_p$  and chemical potential
  $\mu_p$ for the spinless NM-$p$SC junction.  The values of the pairing potential $\Delta_p$ are given in units of $\mu_N/k_F$, while the chemical potential $\mu_p$ and energy $E$ are expressed in units of $\mu_N$. We show typical results
  for the nontopological ($\mu_p<0$, left column), transition
  ($\mu_p=0$, middle column), 
  and topological ($\mu_p>0$, right column) regimes, and for barrier strength
  $Z=0$ (top row) and $Z=1$ (bottom row).}\label{fig:GENp}  
\end{figure}

\begin{figure}[h!]
\capstart
\begin{center}
\includegraphics[width=\linewidth]{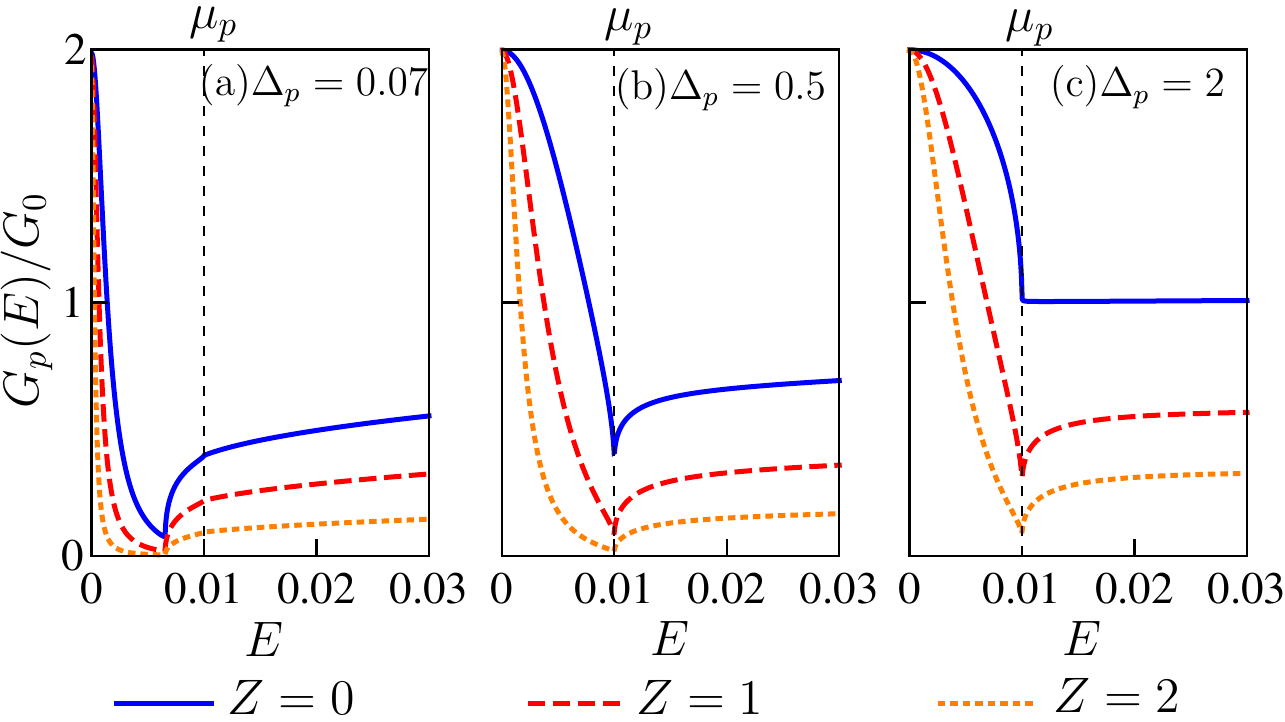}
\end{center}
\caption{(Color online) Variation of the tunneling conductance 
  $G_p(E)$ with the barrier strength $Z$ and the pairing
  potential $\Delta_p$ for the spinless NM-$p$SC junction in the topological regime. The values of the pairing potential $\Delta_p$ are given in units of $\mu_N/k_F$ while the chemical potential $\mu_p$ and energy $E$ are expressed in units of $\mu_N$. Note that the zero-bias
  conductance is constrained to be $2G_0$ by the topological
  condition.}\label{fig:GENpz}  
\end{figure}

Characteristic plots of the conductance as a function of the
energy are shown in
Figs.~\ref{fig:GENp} and~\ref{fig:GENpz}. In general, the tunneling
conductance $G_p(E)$ 
decreases with the barrier strength $Z$, although in the topological
regime the zero-bias conductance is unaffected by $Z$. Furthermore, it is interesting to note that in the topological regime, the width of the zero-bias peak decreases with $Z$ but shows a nonmonotonic dependence with $\Delta_p$; the width first increases as $\Delta_p$ increases, however, beyond a certain value of $\Delta_p$, the width decreases with $\Delta_p$.
For $\mu_p \leq \Delta_p^2/2$, a singularity appears in the $G_p(E)$
curve at the gap edge $E_2 = |\mu_p|$. On the other hand, as shown
in~Fig.~\ref{fig:GENpz}(a), two
singularities are visible in the 
conductance for $\mu_p>\Delta_p^2/2$, corresponding to the
edge of the gap at $E_1
= \Delta_p\sqrt{\mu_p-\Delta_p^2/4}$ 
and the local maximum in the spectrum at $E_2 =
\mu_p$.

\section{Junction with a spin-orbit-coupled nanowire}\label{sec:ns}

In its topological phase, the low-energy sector of the
spin-orbit-coupled nanowire proposal is formally equivalent to the spinless
$p$-wave superconductor studied above~\cite{lutchyn10,oreg10,mourik12}. In order to obtain the
conductance spectrum and its variation across the topological
transition, however, we must examine the full model including spin-orbit coupling and Zeeman splitting. In this section
we therefore consider a one-dimensional junction between a
spin-split spin-orbit-coupled superconducting wire (SOCSW) and a spinful normal
metal (NM), which occupy the regions
$x\geq0$ and $x\leq0$, respectively. Similar to Sec.~\ref{sec:np},
we model their interface at $x=0$ by a $\delta$-potential
barrier of strength $Z$. The Hamiltonian in each region is written $
H_j(x)= \frac{1}{2}\int dx\overline{\Psi}^{\dagger}_j(x)\mathcal{H}_j\overline{\Psi}_j(x)$, where $\overline{\Psi}_j(x) =
(\psi_{j\uparrow}(x),\psi_{j\downarrow}(x),\psi^\dagger_{j\downarrow}(x),-\psi^\dagger_{j\uparrow}(x))^{\mathrm{T}}$
and $\psi^\dagger_{j\sigma}(x)$ [$\psi_{j\sigma}(x)$] is the creation
(annihilation) field operator of an electron with spin $\sigma$ in
region $j=N$ (NM) or $S$ (SOCSW). Using the same unit convention as in
the previous section, we write the BdG Hamiltonians of the NM and 
SOCSW as
\begin{subequations}
\begin{align}
\mathcal{H}_N & = \left(-\partial_x^2- 1 \right)\tau_z, \\
\mathcal{H}_{S} &= -\partial_x^2\tau_z -i \alpha
\partial_x\tau_z\sigma_z + V_Z\sigma_x + \Delta_0\tau_x, \label{eq:HamSOCSW}
\end{align}
\end{subequations}
where $\sigma_{\mu}$ ($\tau_{\mu}$) are the Pauli matrices
in spin (particle-hole) space, $\alpha$ is the strength of
spin-orbit coupling (SOC), $V_Z$ is the Zeeman field, and $\Delta_0 \geq 0$
is the proximity-induced $s$-wave pairing potential which is assumed to be
real. We set the chemical potential of the SOCSW and Zeeman coupling in the lead to be zero,
and take uniform electron masses throughout the system. 

\begin{figure}[t!]
\capstart
\begin{center}
\includegraphics{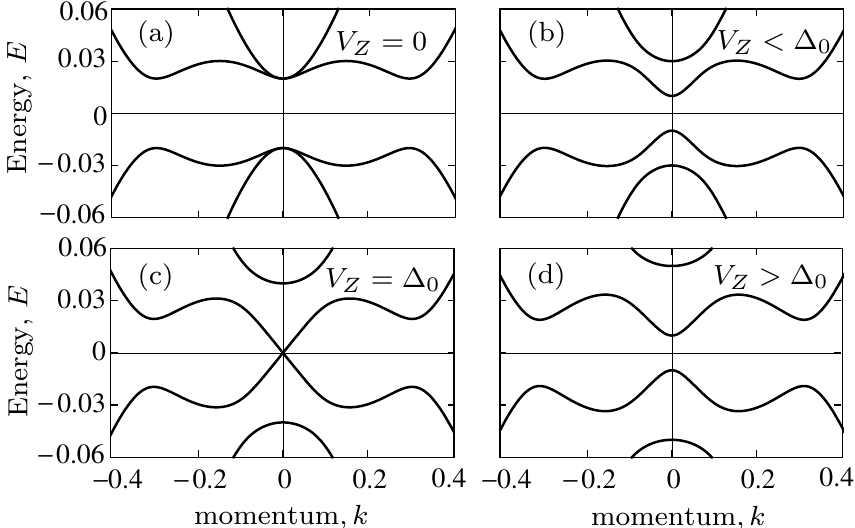}
\end{center}
\caption{Energy spectrum of the SOCSW for different
  values of Zeeman potentials: (a) $V_Z = 0$ (nontopological), (b)
  $V_Z = 0.01$ (nontopological), (c) 
  $V_Z = 0.02$ (transition), and (d) $V_Z = 0.03$
  (topological). In all plots, we set $\alpha = 0.3$ and  $\Delta_0 = 0.02$.} \label{fig:Es}  
\end{figure}

The positive branches of the BdG spectrum of the SOCSW are given by 
\begin{align}\label{eq:Epbar}
E_{\pm} =& \bigg(k^4 + \alpha^2 k^2 + \Delta_0^2 + V_Z^2 \notag \\
&\hspace{1cm} \pm 2 \sqrt{k^4 (\alpha^2
  k^2 + V_Z^2) + \Delta_0^2 V_Z^2}\bigg)^{1/2}. 
\end{align}
As shown in Fig.~\ref{fig:Es}, the energy spectrum is gapped except for
$V_Z = \Delta_0$. This value of $V_Z$ marks the topological quantum phase 
transition between the topologically trivial ($V_Z<\Delta_0$) and
nontrivial phases ($V_Z>\Delta_0$)~\cite{sau10a,lutchyn10,oreg10,sau10b}. 
Although Eq.~\eqref{eq:Epbar} can be analytically solved for the
momenta corresponding to a given energy $E$, the general expression is
unwieldy. In what follows, therefore, we will instead work in the limits of
a strong Zeeman field and strong SOC, where more compact results
can be obtained. 

\subsection{Strong Zeeman splitting}

\begin{figure}
\capstart
\begin{center}
\includegraphics{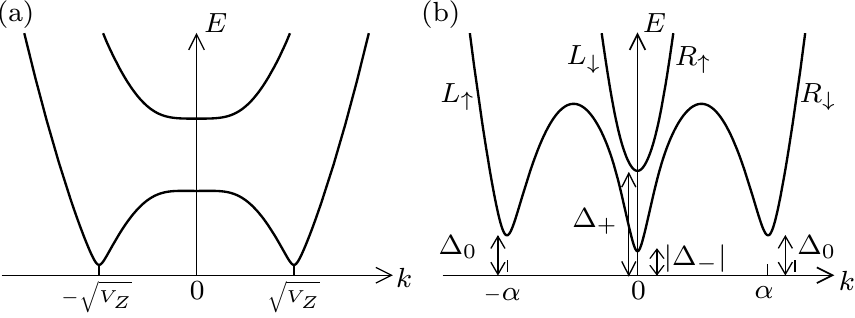}
\end{center}
\caption{Energy spectrum of the SOCSW in the limits
  of (a) strong  Zeeman field and (b) strong SOC. For
  clarity, only the positive energy branches of the spectrum are
  shown. In panel (b), the spectrum about the minima at $k=0$
  constitute the ``interior'' branches, while the spectrum about the
  minima at $k=\pm\alpha$ are the ``exterior''
  branches. Note the different effective gaps for these branches, and
  the states contributing to the slowly varying left- and right-moving
  fields, $L_\sigma(x)$ and $R_\sigma(x)$,
  respectively. } \label{fig:stronglimit}    
\end{figure}

In the limit of strong Zeeman splitting ($V_Z \gg \alpha$, $\Delta_0$), the
quasiparticle excitation spectrum of the SOCSW is split into two spin bands as shown in Fig.~\ref{fig:stronglimit}(a). In the normal state ($\Delta_0=0$) the spectrum is approximately given by 
$\epsilon_\pm(k) \approx k^2 \pm V_Z$. The system is then a
half-metal, with only one spin-polarized band [$\epsilon_-(k)$]
occupied. Projecting the full Hamiltonian into this band gives the
effective Hamiltonian~\cite{lutchyn10,alicea10,sau10b} 
\begin{align}
H_{S}'(k) &= \sum_k\left\{\epsilon_-(k)\psi^{\dagger}_{S-}(k)\psi_{S-}(k) \right.\nonumber\\
&\hspace{1.5 cm}\left.+ \left[\widetilde{\Delta}_-(k)\psi^{\dagger}_{S-}(k)\psi^{\dagger}_{S-}(-k) + \mathrm{h.c.}\right]\right\},
\end{align}
where $\widetilde{\Delta}_-(k) \approx \alpha
k\Delta_0/V_Z$ is a $p$-wave pairing potential and $\psi_{S-}(\psi_{S-}^\dagger)$ is the annihilation (creation) field operator for $\epsilon_-(k)$ band. The projected
Hamiltonian is equivalent to the spinless $p$SC Hamiltonian $H_{p}(k)$
[Eq.~\eqref{eq:hnp}], with the identifications $\mu_p
= V_Z$ and $\Delta_p = \alpha\Delta_0/V_Z$. If the
Zeeman field is applied on both sides of the junction such that the NM
is also fully spin polarized, then the low-energy sector is identical
to the spinless  NM-$p$SC junction studied in Sec.~\ref{sec:np}, and
the results obtained above for the differential conductance directly  
apply.  

\subsection{Strong spin-orbit coupling}

In the case of strong SOC ($\alpha \gg V_Z,\Delta_0$), the BdG
spectrum of the SOCSW has the characteristic form shown
in~\fig{fig:stronglimit}(b). In particular, we note that both the $+$
and $-$ spectra [Eq.~\eqref{eq:Epbar}] have minima at $k=0$ (the so-called interior branches),
while the $-$ spectrum also has minima at $k=\pm\alpha$ (the exterior
branches). For small energies $E\lesssim \Delta_0$, $V_Z$, we can
linearize the Hamiltonian about these minima by introducing the ansatz
for the field operators~\cite{braunecker10,klinovaja12a,klinovaja12b}
\begin{subequations}
\begin{align}
\psi_{S\uparrow}(x) &\approx R_{\uparrow}(x) + L_{\uparrow}(x)e^{-i\alpha x}, \\
\psi_{S\downarrow}(x) &\approx L_{\downarrow}(x) + R_{\downarrow}(x)e^{i\alpha x},
\end{align}
\end{subequations}
where $R_{\sigma}(x)$ and $L_{\sigma}(x)$ represent slowly-varying right-
and left-moving fields, respectively;
see~\fig{fig:stronglimit}(b). Inserting this ansatz into the 
Hamiltonian [\eq{eq:HamSOCSW}] and neglecting all ``fast
oscillating'' terms (involving terms with phase
factors $e^{\pm i\alpha x}$), we obtain effective Hamiltonians valid
for the states near the interior and exterior branches. Specifically,
we write
\beq
\widetilde{H}_{S}^{(l)} = \frac{1}{2}\int dx
\tp^{(l)}_S(x)^{\dagger}\mathcal{\widetilde{H}}_{S}^{(l)}\tp^{(l)}_S(x),
\eeq
where $l= e$, $i$ denotes the exterior and interior branches,
respectively, and the BdG Hamiltonians are written
\begin{subequations}\label{eq:hie}
\begin{align}
\widetilde{\mathcal{H}}_{S}^{(e)} & = -i\alpha\tau_z\sigma_z\partial_x +\Delta_0\tau_x\,, \\
\widetilde{\mathcal{H}}_{S}^{(i)} & = -i\alpha\tau_z\sigma_z\partial_x +V_Z\sigma_x +\Delta_0\tau_x\,.
\end{align}
\end{subequations}
The spinors for the interior and exterior branches are defined in
terms of the slowly-varying field as
$\widetilde{\Psi}_S^{(e)}(x) = (L_{\uparrow}(x),R_{\downarrow}(x), 
R_{\downarrow}^{\dagger}(x),-L_{\uparrow}^{\dagger}(x))^{\mathrm{T}}$ and
$\widetilde{\Psi}_S^{(i)}(x) =
(R_{\uparrow}(x),L_{\downarrow}(x),L_{\downarrow}^{\dagger}(x),-R_{\uparrow}^{\dagger}(x))^{\mathrm{T}}$. 

We consider an electron with energy $E$ and spin $\sigma$ injected
into the SOCSW from the NM. The wave function in the NM is given by 
\beq
\Phi_{N\sigma}(x)
= \left(\begin{array}{c}
\delta_{\sigma\uparrow}\\
\delta_{\sigma\downarrow}\\
0\\
0\\
\end{array}\right) e^{ix} + \left(\begin{array}{c}
b_{\sigma\uparrow}\\
b_{\sigma\downarrow}\\
0\\
0\\
\end{array}\right)e^{-ix} +
\left(\begin{array}{c}
0\\
0\\
a_{\sigma\downarrow}\\
a_{\sigma\uparrow}\\
\end{array}\right)e^{ix}\,,
\eeq
 where $\delta_{\sigma\sigma'}$ is the Kronecker symbol. The
 coefficients $a_{\sigma\sigma'}$ and 
 $b_{\sigma\sigma'}$ are the amplitudes for Andreev and normal   
 reflection, respectively. Note that due to the SOC in
 the SOCSW, both spin-flip and spin-preserving reflection processes
 are allowed. The wave function in the SOCSW is a superposition of
 solutions on the exterior and interior branches
\begin{align}
\lefteqn{\Phi_{S\sigma}(x) =} \notag \\
& c^{(i)}_{\sigma 1}\left(\begin{array}{c}
-u_- \\
\sgn(\Delta_-)v_- \\
-\sgn(\Delta_-)v_- \\
u_-
\end{array}\right)e^{ik_-^{(i)}x} +
c^{(i)}_{\sigma 2}\left(
\begin{matrix}
u_+ \\
v_+ \\
v_+ \\
u_+
\end{matrix}
\right) e^{ik_+^{(i)}x} \notag \\
& + c^{(e)}_{\sigma 1}\left(
\begin{matrix}
v_0 \\
0 \\
u_0\\
0
\end{matrix} \right)e^{i( k_0^{(e)}-\alpha)x} + c^{(e)}_{\sigma 2}\left(\begin{matrix}
0  \\
u_0\\
0\\
v_0
\end{matrix} \right)e^{i( k_0^{(e)}+\alpha)x}\,, \label{eq:PsiSOCSW}
\end{align}
where the first line on the right-hand side gives contributions from the interior
branches, while the second line originates from the exterior
branches. Note that the coefficients
$c^{(i)}_{\sigma(1,2)}$ and $c^{(e)}_{\sigma(1,2)}$ are the transmission coefficients into the SOCSW. The elements of the wave function are given by
\begin{align}
u_{\nu}^2 &=
\begin{cases}
\left(E + \sqrt{E^2 - \Delta_{\nu}^2}\right)/2E, & \mathrm{for} \hspace{0.2 cm} E \geq
|\Delta_{\nu}|, \\
\left(E + i\sqrt{\Delta_{\nu}^2-E^2}\right)/2|\Delta_\nu|, & \mathrm{for} \hspace{0.2 cm}  0\leq E < |\Delta_{\nu}|, 
\end{cases}
\end{align}
and
\begin{align}
v^2_\nu +u_\nu^2 & = \begin{cases}
1, & \mathrm{for} \hspace{0.2 cm} E \geq |\Delta_\nu|, \\
E/|\Delta_\nu|, & \mathrm{for} \hspace{0.2 cm} 0\leq E < |\Delta_\nu|,
\end{cases}
\end{align}
where $\nu=\pm$, $0$, and $\Delta_\pm = \Delta_0 \pm V_Z$. The wave vectors
appearing in~\eq{eq:PsiSOCSW} are $k_{\pm}^{(i)} = \sqrt{E^2 - 
  \Delta_{\pm}^2}/\alpha$ for the interior branches, and $k_0^{(e)} = \sqrt{E^2 -
  \Delta_{0}^2}/\alpha$ for the exterior branches.

The wave functions satisfy the continuity and current conservation
boundary conditions 
\begin{subequations}\label{eq:boundns}
\begin{gather}
\Phi_{S\sigma}(x)|_{x=0^+} = \Phi_{N\sigma}(x)|_{x=0^-}\,,\\
J_{S}\Phi_{S\sigma}(x)|_{x=0^+} - J_{N}\Phi_{N\sigma}(x)|_{x=0^-} = -2iZ\tau_z\Phi_{N\sigma}(0),
\end{gather}
\end{subequations}
where the current operators are given by
\beq
J_{N} = -2i\partial_x\tau_z\,, \qquad
J_{S} = -2i\partial_x\tau_z + \alpha\tau_z\sigma_z\,.
\eeq
In the limit of strong SOC ($\alpha \gg V_Z,\Delta_0$), we ignore
terms proportional to $k_-^{(i)}$, $k_+^{(i)}$, $k_0^{(e)} \ll1$ in the current
conservation equation. Expressions for the Andreev and normal
reflection coefficients found from solving these equations are given
in Appendix~\ref{ap:ns}. 

\begin{table*}[htb]
\capstart
\setlength{\tabcolsep}{10pt}
\renewcommand{\arraystretch}{2.6}
\begin{tabular}{|c|c|c|c|}
\hline
& $V_Z < \Delta_0$ & $V_Z = \Delta_0$ & $V_Z > \Delta_0$ \\\hline
$a_{\uparrow\uparrow}(0)$ & $0$ & $\displaystyle{\frac{\alpha[1+(Z+i\alpha/2)^2]}{D_1D_2}}$ & $\displaystyle{\frac{1+(Z+i\alpha/2)^2}{D_1}}$ \\\hline
$a_{\uparrow\downarrow}(0)$ & $\displaystyle{-\frac{2i\alpha}{D_1}}$ & $\displaystyle{-\frac{i\alpha}{D_1}}$ & $\displaystyle{\frac{i}{2}-\frac{i\alpha}{D_1}}$  \\\hline
$a_{\downarrow\uparrow}(0)$ & $\displaystyle{-\frac{2i\alpha}{D_1}}$ &
$\displaystyle{-i\alpha\left(\frac{1}{D_1}+\frac{1}{D_2}\right)}$ &
$\displaystyle{-\frac{i}{2}-\frac{i\alpha}{D_1}}$  \\\hline 
$a_{\downarrow\downarrow}(0)$ & $0$ &
$\displaystyle{\frac{\alpha[1+(Z-i\alpha/2)^2]}{D_1D_2}}$ &
$\displaystyle{\frac{1+(Z-i\alpha/2)^2}{D_1}}$ \\\hline 
$b_{\uparrow\uparrow}(0)$ & $\displaystyle{\frac{2[(i+Z)^2+(\alpha/2)^2]}{D_1}}$ &  $\displaystyle{\frac{2[(i+Z)^2+(\alpha/2)^2][D_2-\alpha/2]}{D_1D_2}}$ & $\displaystyle{\frac{(i+Z)^2+(\alpha/2)^2}{D_1}}$\\\hline
$b_{\uparrow\downarrow}(0)$ & $0$ &
$\displaystyle{\frac{-i\alpha(1-iZ+\alpha/2)^2}{D_1D_2}}$ &
$\displaystyle{\frac{-i(1-iZ+\alpha/2)^2}{D_1}}$\\\hline 
$b_{\downarrow\uparrow}(0)$ & $0$ & $\displaystyle{\frac{i\alpha(-1+iZ+\alpha/2)^2}{D_1D_2}}$
& $\displaystyle{\frac{i(-1+iZ+\alpha/2)^2}{D_1}}$\\\hline 
$b_{\downarrow\downarrow}(0)$ & $\displaystyle{\frac{2[(i+Z)^2+(\alpha/2)^2]}{D_1}}$ &  $\displaystyle{\frac{2[(i+Z)^2+(\alpha/2)^2][D_2-\alpha/2]}{D_1D_2}}$ & $\displaystyle{\frac{(i+Z)^2+(\alpha/2)^2}{D_1}}$\\\hline
$\displaystyle{\frac{G_S(0)}{G_0}}$ & $\displaystyle{\frac{16\alpha^2}{D_1^2}}$ &
$\displaystyle{2\alpha\left(\frac{4}{D_1}-\frac{1}{D_2}\right)}$ & $2$\\\hline 
\end{tabular}
\caption{Zero-bias values of the Andreev reflection coefficients
  $a_{\sigma\sigma'}(0)$,  normal reflection coefficients
  $b_{\sigma\sigma'}(0)$, and differential conductance $G_S(0)$ in the
  strong SOC limit of the NM-SOCSW junction. The three
  columns give the values in the nontopological ($V_Z<\Delta_0$),
  transition ($V_Z=\Delta_0$) and topological ($V_Z>\Delta_0$) 
  regimes. The terms $D_{1,2}$ are given by $D_1 =
  2[1+Z^2+(\alpha/2)^2]$ and $D_2 =Z^2+(1+\alpha/2)^2$.}\label{tab:ns} 
\end{table*}

\begin{figure}[h!]
\capstart
\begin{center}
\includegraphics[width=\linewidth]{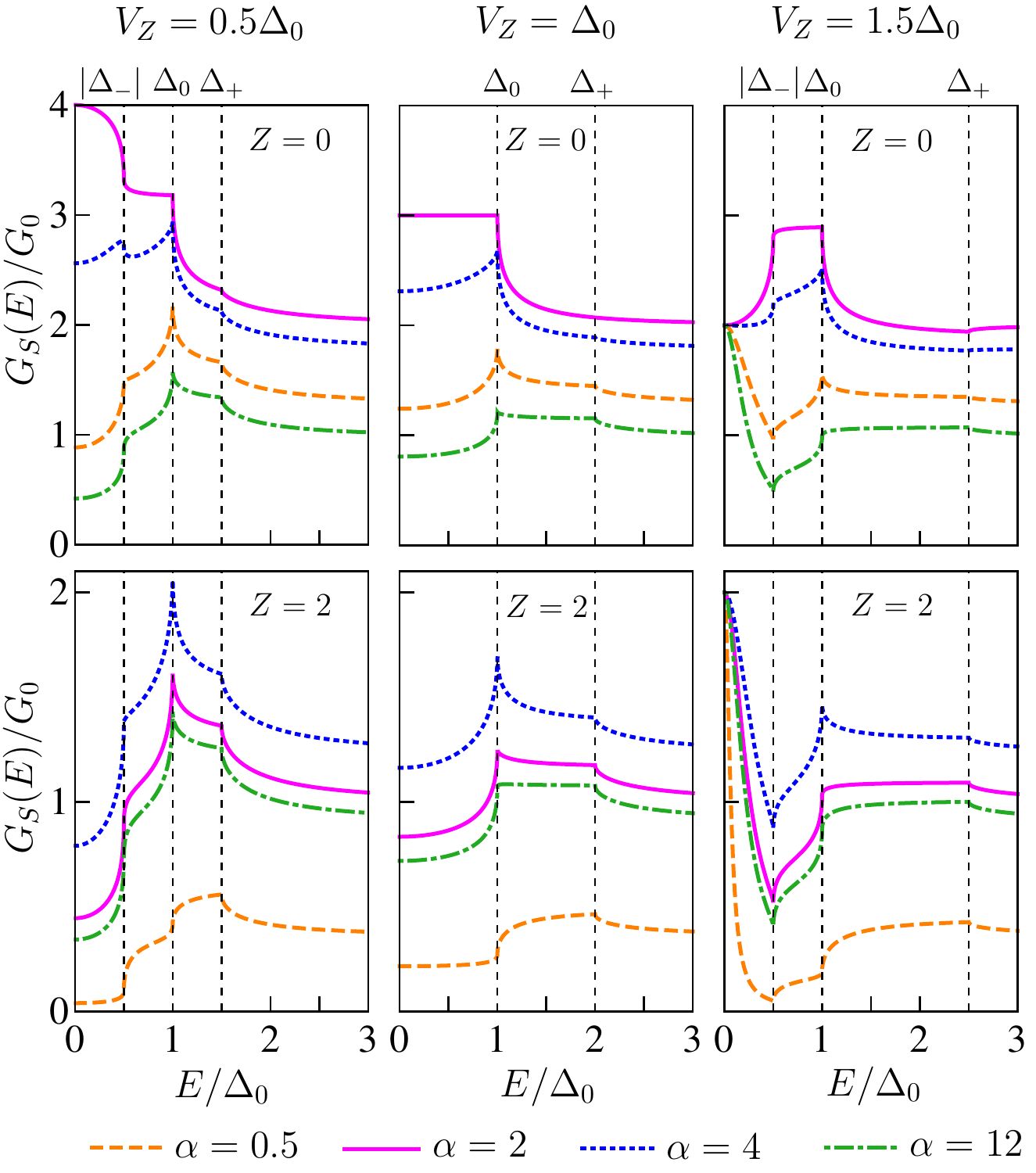}
\end{center}
\caption{(Color online) Variation of the tunneling conductance 
  $G_S(E)$ with SOC strength $\alpha$ and Zeeman field $V_Z$ in the
  strong SOC limit of the NM-SOCSW junction. We
  present typical results for the nontopological ($V_Z<\Delta_0$, left
  column), 
  transition ($V_Z=0$, middle column), and topological
  ($V_Z>\Delta_0$, right column) regimes, 
  and for barrier strength $Z=0$ (top row) and $Z=2$ (bottom row). In
  all plots we set $\Delta_0 = 0.001$. The values of $\Delta_0$ and $V_Z$ are given in units of $\mu_N$, while the values of $\alpha$ are expressed in units of $\mu_N/k_F$.}\label{fig:GENS} 
\end{figure}

\begin{figure}[h!]
\capstart
\begin{center}
\includegraphics[width=\linewidth]{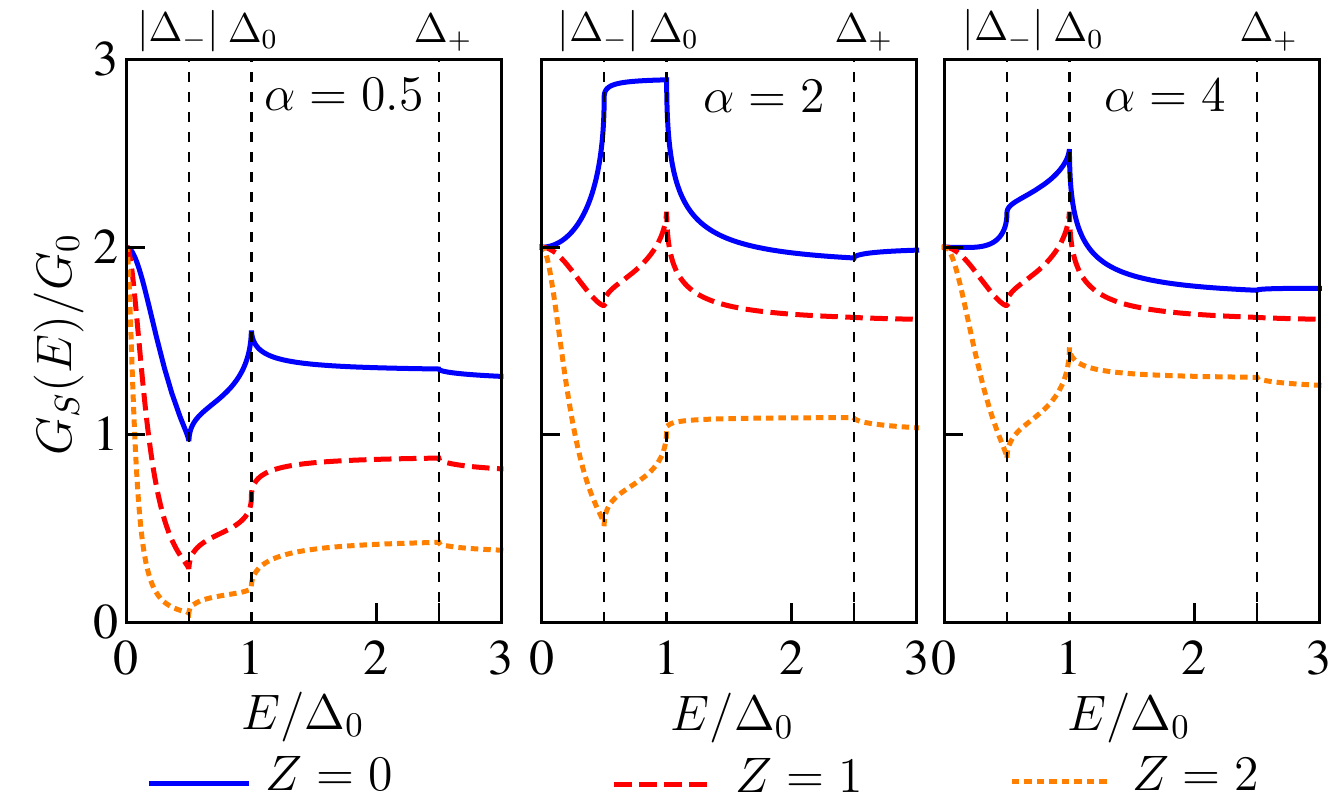}
\end{center}
\caption{(Color online) Variation of the tunneling conductance 
  $G_S(E)$ with interface barrier strength $Z$ and spin-orbit coupling
  strength $\alpha$ for the strong SOC limit of the NM-SOCSW junction in the
  topological regime. We set  
  $\Delta_0 = 0.001$ and $V_Z = 1.5\Delta_0$. The values of $\Delta_0$ and $V_Z$ are given in units of $\mu_N$, while the values of $\alpha$ are expressed in units of $\mu_N/k_F$. Note that in all cases
  the zero-bias conductance is equal to $2G_0$, consistent with the
  topological state.}\label{fig:GENSz} 
\end{figure}

The zero-temperature
differential tunneling conductance ${G}_{S}(E)$ is obtained from the
BTK formula   
\begin{align}\label{eq:Gsemiconductor}
G_{S}(E) &= G_0 \left( 2 + \sum_{\sigma,\xi=\uparrow,\downarrow}\left\{|a_{\sigma\xi}(E)|^2 - |b_{\sigma\xi}(E)|^2\right\}\right)\,.
\end{align}
Although the general expression is complicated, compact forms for the
reflection coefficients and the conductance at zero bias are presented
in Table~\ref{tab:ns}. As in the spinless NM-$p$SC junction studied
above, the zero-bias conductance $G_S(0)$ is discontinuous across the
topological phase transition. In the topological regime
($V_Z>\Delta_0$) the zero-bias conductance takes the quantized value
$G_S(0) = 2$. This implies that the Andreev reflection
coefficients in~\eq{eq:Gsemiconductor} exactly cancel the normal
reflection coefficients; moreover, from Table~\ref{tab:ns} it can be
verified that $\sum_{\sigma,\xi}|a_{\sigma\xi}(0)|^2 =
\sum_{\sigma,\xi}|b_{\sigma\xi}(0)|^2 = 1$. This can be understood in
terms of the existence of a single Majorana mode at the interface
which couples to one of the two channels in the normal
region~\cite{Law09,Xin15}. While 
there is perfect Andreev reflection in this channel, in the other channel
we have perfect normal reflection. 
In the nontopological regime, on the other hand,
$G_S(0)$ takes on nonuniversal values and is dependent upon $Z$ and
$\alpha$. In particular, the zero-bias conductance in the nontopological phase can strongly exceed
the quantized value in the topological state: for the gapped nontopological state
($V_Z<\Delta_0$) and at the topological transition point
($V_Z=\Delta_0$), we find the maximum values $G_S(0)=4$ and  $G_S(0) =
3$, respectively, which are realized for $Z =0$ and
$\alpha=2$. 

We plot the calculated conductance as a function of energy in
Figs.~\ref{fig:GENS} and~\ref{fig:GENSz}. In the former we show
examples of the conductance spectra in the nontopological, transition,
and topological regimes, while the latter explores more fully the
variation of the conductance spectra in the topological state away from zero bias.  
The conductance spectra show a much more complicated structure than those in
the spinless NM-$p$SC junction, reflecting the presence of three
distinct gaps ($\Delta_+$, $|\Delta_-|$, $\Delta_0$) in the strong SOC
limit of the SOCSW. Indeed, at the energy corresponding to each gap we
observe a nonanalyticity in the conductance spectrum. Although there is
considerable variation in the conductance spectrum as a function of energy, a
number of trends can be discerned: increasing $Z$ tends to suppress the
conductance, the energy variation of the conductance is nonmonotonic
in general with cusplike structures at specific energies,
and the energy variation of the conductance is stronger near
  zero energy for larger values of $Z$. While the conductance at
first 
tends to be enhanced by increasing the SOC, the conductance eventually
goes through a maximum before monotonically decreasing. Similarly, the
SOC increases the width of the zero-bias peak in the topological
regime, but beyond a certain SOC strength it decreases again. The
basic finding is that, other than the universal quantized Majorana
peak at zero energy, the tunneling conductance shows interesting and
nontrivial dependence on $Z$ and $E$ in the topological phase. In
particular, an interesting conclusion of our theory is that the
zero-bias conductance could be quantized in the topological phase for
small values of $Z$ without developing a peak in the tunneling
conductance at all. 

Note that the above discussion holds true also for the case where the Zeeman coupling in the lead or the chemical potential $\mu_S$ of the SOCSW are nonzero. For the case where $|\Delta_-|<\Delta_0$, the zero-bias peak formed in the topological regime is within an energy range of $|\Delta_-|$. Since the topological gap $|\Delta_-| = |\sqrt{\Delta_0^2+\mu_S^2}-V_Z|$ decreases with the absolute value of the chemical potential $|\mu_S|$, the width of the zero-bias peak decreases with $|\mu_S|$.

\section{Summary}\label{sec:summary}
Using the BTK formalism we have analytically studied the zero-temperature tunneling
conductance spectra of NS junctions involving topological
superconductors. Finite temperature effects within this formalism simply lead to thermal broadening of the zero-temperature conductance and can be included in the theory numerically by introducing an integration over the Fermi function~\cite{eugene}.  As in the BTK paper~\cite{BTK}, the finite-voltage conductances are found to depend on the strength of the barrier at the interface, which 
is parameterized by the dimensionless parameter $Z$. Specifically, we have examined a spinless NM-$p$SC
junction and a spinful NM-SOCSW junction, paying particular attention
to the change in the zero-bias conductance across the topological
phase transition. We explicitly demonstrate that the zero-temperature
zero-bias conductance is quantized at a value of $2e^2/h$ in the topological
regime, in agreement with effective models of these systems based on a
single Majorana mode coupled to a normal channel. Despite
  this quantization at zero voltage, the zero-bias conductance only
  develops a peak (or a local maximum) as a function of voltage for
  barriers with sufficiently large $Z$ parameter, or for small and
  large SOC strength. These parameters also control the width of this
  peak. In the 
nontopological regime, on the other hand, the conductance takes
nonuniversal values depending upon the details of the system. In both
cases the conductance spectrum away from zero bias shows considerable
variation with the details of the junction. Our calculated BTK conductance also shows that the conductance is finite inside the superconducting gap region because of the finite barrier transparency, providing a possible mechanism for the observed ``soft gap" feature in the experimental studies~\cite{mourik12,deng12,das12,Finck2013,churchill13}.  This effect is qualitatively similar to the ``inverse proximity effect" at the NS interface arising from the finite barrier at the interface as discussed in the recent literature~\cite{Stanescu14}, although other possible physical mechanisms for the soft gap behavior have also been proposed~\cite{takei}.  We mention finally that our theory is for a single NS junction which effectively assumes the existence of only a single Majorana mode at the NS interface (with the other Majorana being located infinitely far away) and thus Majorana splitting~\cite{cheng09,cheng10,dassarma12,Rainis13} due to the wave function overlap between two Majorana modes is not germane to our theory (but can be included if necessary in a future generalization).

Finally, we emphasize that one of the most salient features of our theoretical work is that it is completely analytical within a continuum model in contrast to most theoretical works on Majorana nanowires which focus on numerical simulations within a tight-binding lattice model.  All the microscopic details of the complex normal-superconductor tunneling process are simply subsumed in a single phenomenological parameter $Z$ (``the interface barrier strength" of the BTK formalism) allowing our theory a great deal of flexibility for actual modeling of the Majorana nanowire experimental results, since the realistic microscopic details of the NS interface are rarely known in the actual experimental nanowire setups.  It is gratifying that our analytical model captures the essential features of the Majorana nanowire experiments through our finding of the Majorana zero-bias conductance quantization and soft gap feature, with the interesting prediction that for strong metallic junction (i.e., for very low interface barrier or a small value of $Z$) the Majorana zero-bias conductance may not necessarily be a peak in the tunneling conductance although it would still be quantized since the Majorana zero mode necessarily implies perfect Andreev reflection.

\section{Acknowledgments}
We thank H.-Y.~Hui, X.~Liu, and D.~Rainis for fruitful
discussions. This work was supported by Microsoft Station Q, LPS-CMTC,
and JQI-NSF-PFC.
 
\begin{widetext}
\appendix

\section{Andreev and normal reflection coefficients for the NM-SOCSW
  junction}~\label{ap:ns} 
Solving the boundary equations, we obtain the Andreev (normal)
reflection coefficients $a_{\sigma\sigma'}$ ($b_{\sigma\sigma'}$) as 
\begin{eqnarray}
a_{\uparrow\uparrow}(E) &=& -\frac{\alpha u_0v_0[u_-v_+ - \sgn(\Delta_-)u_+v_-][1+(Z+i\frac{\alpha}{2})^2]}{\mathcal{D}_{E}^{(1)}\mathcal{D}_{E}^{(2)}},\\
a_{\uparrow\downarrow}(E) &=& \alpha u_0 \left[\frac{v_+}{\mathcal{D}_E^{(1)}} + \sgn(\Delta_-)\frac{v_-}{\mathcal{D}_{E}^{(2)}}\right], \\
a_{\downarrow\uparrow}(E) &=&\alpha v_0\left(\frac{u_+}{\mathcal{D}_{E}^{(1)}} + \frac{u_-}{\mathcal{D}_{E}^{(2)}}\right),\\
a_{\downarrow\downarrow}(E) &=& -\frac{\alpha u_0v_0\left[u_-v_+-\sgn(\Delta_-)u_+v_-\right]\left[1+(Z-i\frac{\alpha}{2})^2\right]}{\mathcal{D}_{E}^{(1)}\mathcal{D}_{E}^{(2)}},\\
b_{\uparrow\uparrow}(E)  &=& b_{\downarrow\downarrow}(E)\\
&=&
-\left[(i+Z)^2+\left(\frac{\alpha}{2}\right)^2\right]\nonumber\times\\
&&\left\{\frac{\sgn(\Delta_-)v_0^2v_-v_+[Z^2+(\frac{\alpha}{2}-1)^2]-u_0v_0[u_-v_++\sgn(\Delta_-)u_+v_-][1+Z^2+(\frac{\alpha}{2})^2] + u_0^2u_-u_+[Z^2+(1+\frac{\alpha}{2})^2]}{\mathcal{D}_{E}^{(1)}\mathcal{D}_{E}^{(2)}}\right\},\nonumber\\
b_{\uparrow\downarrow}(E) &=&\frac{\alpha
u_0^2[u_-v_+-\sgn(\Delta_-)u_+v_-](1-iZ+\frac{\alpha}{2})^2}{\mathcal{D}_{E}^{(1)}\mathcal{D}_{E}^{(2)}},\\ 
b_{\downarrow\uparrow}(E) &=& \frac{\alpha
v_0^2[u_-v_+-\sgn(\Delta_-)u_+v_-](-1+iZ+\frac{\alpha}{2})^2}{\mathcal{D}_{E}^{(1)}\mathcal{D}_{E}^{(2)}}, 
\end{eqnarray}
where $\mathcal{D}_{E}^{(1)}=u_0u_+[Z^2+(\alpha/2+1)^2] - v_0v_+[Z^2+(\alpha/2-1)^2]$ and $\mathcal{D}_{E}^{(2)} = u_0u_-[Z^2+(\alpha/2+1)^2]-\sgn(\Delta_-)v_0v_-[Z^2+(\alpha/2-1)^2]$. 
\end{widetext}


\begin{thebibliography}{99}

\bibitem{ReadGreen2000}N. Read and D. Green, Phys. Rev. B {\bf 61},
  10267 (2000).

\bibitem{kitaev2001} A. Kitaev, Phys. Usp. \textbf{44}, 131 (2001).

\bibitem{NayakRMP2008}C. Nayak, S. H. Simon, A. Stern, M. Freedman,
  and S. Das Sarma, Rev. Mod. Phys. {\bf 80}, 1083 (2008).

\bibitem{sau10a} Jay D.~Sau, R.~M.~Lutchyn, S.~Tewari, and S.~Das
  Sarma, Phys. Rev. Lett. \textbf{104}, 040502 (2010). 

\bibitem{sau10b} Jay D.~Sau, S.~Tewari, R.~M.~Lutchyn, T.~D.~Stanescu and S.~Das
  Sarma, Phys. Rev. B \textbf{82}, 214509 (2010).

\bibitem{lutchyn10} R.~M.~Lutchyn, Jay D. Sau, and S. Das Sarma,
  Phys. Rev. Lett. \textbf{105}, 077001 (2010).

\bibitem{oreg10} Y. Oreg, G. Refael, and F. von Oppen,
  Phys. Rev. Lett. \textbf{105}, 177002 (2010).

\bibitem{alicea10} J.~Alicea, Phys. Rev. B \textbf{81}, 125318 (2010).

\bibitem{mourik12}  V. Mourik, K. Zuo, S. M. Frolov, S. R. Plissard,
  E. P. A. M. Bakkers, and L. P. Kouwenhoven, Science \textbf{336},
  1003 (2012).

\bibitem{Rokhinson12} L. P. Rokhinson, X. Liu, and J. K. Furdyna,
  Nat. Phys. {\bf 8}, 795 (2012).

\bibitem{deng12} M.~T.~Deng, C.~L.~Yu, G.~Y.~Huang, M.~Larrson,
  P.~Caroff, and H.~Q.~Xu, Nano Lett. \textbf{12}, 6414 (2012).

\bibitem{das12} A.~Das, Y.~Ronen, Y.~Most, Y.~Oreg, M.~Heiblum, and
  H.~Shtrikman, Nat. Phys. \textbf{8}, 887 (2012).

\bibitem{Finck2013} A. D. K. Finck, D. J. Van Harlingen,
  P. K. Mohseni, K. Jung, and X. Li, Phys. Rev. Lett. {\bf 110},
  126406 (2013).

\bibitem{churchill13}H. O. H. Churchill, V. Fatemi,
  K. Grove-Rasmussen, M. T. Deng, P. Caroff, H. Q. Xu, and
  C. M. Marcus, Phys. Rev. B {\bf 87}, 241401(R) (2013).

\bibitem{Lee13} E. J. H. Lee, X. Jiang, M. Houzet, R. Aguado,
  C. M. Lieber, and S. De Franceschi, Nat. Nanotechnol. {\bf 9}, 79
  (2014).

\bibitem{Fu08} L. Fu and C. L. Kane, Phys. Rev. Lett. \textbf{100}, 096407 (2008).

\bibitem{Fu09} L. Fu and C. L. Kane, Phys. Rev. B \textbf{79}, 161408(R) (2009).

\bibitem{Choy11} T.-P. Choy, J. M. Edge, A. R. Akhmerov, and C. W. J. Beenakker, Phys. Rev. B \textbf{84}, 195442 (2011).

\bibitem{Mi13} S.~Mi, D.~I.~Pikulin, M.~Wimmer, and C.~W.~J.~Beenakker Phys. Rev. B \textbf{87}, 241405(R) (2013). 

\bibitem{Zhang08} C. Zhang, S. Tewari, R. M. Lutchyn, and S. Das Sarma, Phys. Rev. Lett. \textbf{101}, 160401 (2008).

\bibitem{Sato09} M. Sato, Y. Takahashi, and S. Fujimoto, Phys. Rev. Lett. \textbf{103}, 020401 (2009).

\bibitem{Duckheim11} M. Duckheim and P. W. Brouwer, Phys. Rev. B \textbf{83}, 054513 (2011).

\bibitem{Chung11} S. B. Chung, H.-J. Zhang, X.-L. Qi and S.-C. Zhang, Phys. Rev. B \textbf{84}, 060510(R) (2011).

\bibitem{Takei11} S. Takei and V. Galitski, Phys. Rev. B \textbf{86}, 054521 (2012).

\bibitem{Mao12} L. Mao, M. Gong, E. Dumitrescu, S. Tewari, and C. Zhang, Phys. Rev. Lett. \textbf{108}, 177001 (2012). 

\bibitem{Sau12} Jay D.~Sau and S.~Das Sarma, Nat. Commun. \textbf{3}, 964 (2012).

\bibitem{Kim14} Y.~Kim, M.~Cheng, B.~Bauer, R.~M.~Lutchyn, and S.~Das Sarma, Phys. Rev. B \textbf{90}, 060401(R) (2014).

\bibitem{Brydon15} P.~M.~R.~Brydon, S.~Das Sarma, H.-Y. Hui, and Jay D. Sau, Phys. Rev. B \textbf{91}, 064505 (2015).

\bibitem{Hoi15} H.-Y.~Hui, P.~M.~R.~Brydon, Jay D. Sau and S.~Das Sarma, Sci. Rep. \textbf{5}, 8880 (2015).

\bibitem{Alicea12} J. Alicea, Rep. Prog. Phys. \textbf{75}, 076501 (2012).
\bibitem{DasSarma15} S.~Das Sarma, M.~Freedman, and C.~Nayak, arXiv:1501.02813.
\bibitem{Tudor13} T. D. Stanescu and S. Tewari, J. Phys. Condens. Matter \textbf{25}, 233201 (2013).
\bibitem{Franz15} S. R. Elliott and M. Franz, Rev. Mod. Phys. \textbf{87}, 137 (2015).

\bibitem{Sengupta01} K.~Sengupta,  I. \u{Z}uti\'{c}, H.-J. Kwon,
  V. M. Yakovenko, and S. Das Sarma, Phys. Rev. B \textbf{63}, 144531
  (2001).

\bibitem{Law09} K.~T.~Law, P.~A.~Lee, and T.~K.~Ng,
  Phys. Rev. Lett. \textbf{103}, 237001 (2009).

\bibitem{Flensberg10} K. Flensberg, Phys. Rev. B \textbf{82}, 180516
  (2010).

\bibitem{Wimmer11} M.~Wimmer, A.~R.~Akhmerov, J.~P.~Dahlhaus, and
  C.~W.~J.~Beenakker, New. J. Phys. \textbf{13}, 053016 (2011).
	
\bibitem{Xin15} X.~Liu, Jay D. Sau, and S. Das Sarma, arxiv:1501.07273.

\bibitem{Lin12} C.-H.~Lin, Jay D.~Sau and S. Das Sarma, Phys. Rev. B
  \textbf {86}, 224511 (2012).
	
\bibitem{Pientka12} F. Pientka, G. Kells, A. Romito, P. W. Brouwer, and F. von Oppen, Phys. Rev. Lett. \textbf{109}, 227006 (2012).
	
\bibitem{Kells12}G. Kells, D. Meidan, and P. W. Brouwer, Phys. Rev. B
  {\bf 86}, 100503(R) (2012).

\bibitem{Liu12}J. Liu, A. C. Potter, K. T. Law, and P. A. Lee,
  Phys. Rev. Lett. {\bf 109}, 267002 (2012).

\bibitem{Bagrets12}D. Bagrets and A. Altland, Phys. Rev. Lett. {\bf
  109}, 227005 (2012).

\bibitem{Pikulin12}D. I. Pikulin, J. P. Dahlhaus, M. Wimmer,
  H. Schomerus, and C. W. J.  Beenakker, New. J. Phys. {\bf 14},
  125011 (2012). 

\bibitem{Rainis13}D. Rainis, L. Trifunovic, J. Klinovaja, and D. Loss,
  Phys. Rev. B {\bf 87}, 024515 (2013).

\bibitem{Roy13}D. Roy, N. Bondyopadhaya, and S. Tewari, Phys. Rev. B
  {\bf 88}, 020502(R) (2013).
	

	
\bibitem{Qu11}C. Qu, Y. Zhang, L. Mao, and C. Zhang, arXiv:1109.4108
  (unpublished). 

\bibitem{Stanescu11}T. D. Stanescu, R. M. Lutchyn, and S. Das Sarma,
  Phys. Rev. B {\bf 84}, 144522 (2011). 

\bibitem{Prada12} E. Prada, P. San-Jose, and R. Aguado, Phys. Rev. B \textbf{86}, 180503(R) (2012).

\bibitem{Stanescu14} T. D. Stanescu, R. M. Lutchyn, and S. Das Sarma, Phys. Rev. B {\bf 90}, 085302 (2014). 

\bibitem{Dibyendu12} D. Roy, C. J. Bolech, and N. Shah, Phys. Rev. B {\bf 86}, 094503 (2012).

\bibitem{Dibyendu13} D. Roy, C. J. Bolech, and N. Shah, arxiv:1303.7036.

\bibitem{James14} J.~J.~He, T.~K.~Ng, P.~A.~Lee, and K.~T.~Law,
  Phys. Rev. Lett. \textbf{112}, 037001 (2014).

\bibitem{Yan14a} Z.~Yan and S.~Wan, New J. Phys. \textbf{16} 093004
  (2014).

\bibitem{Rex14}S. Rex and A.~Sudb\o{}, Phys. Rev. B {\bf 90}, 115429 (2014).

\bibitem{Thakurathi12} M.~Thakurathi, O.~Deb and D.~Sen,
  arxiv:1412.0072.

\bibitem{Yan14b} Z.~Yan and S.~Wan, arxiv:1411.5919v2.

\bibitem{Sun15} K.~Sun and N.~Shah, Phys. Rev. B {\bf 91}, 144508 (2015).

\bibitem{BTK} G.~E.~Blonder, M.~Tinkham and T.~M.~Klapwijk,
  Phys. Rev. B \textbf{25}, 4515 (1982).
	
\bibitem{Tanaka94} Y.~Tanaka and S.~Kashiwaya, Phys. Rev. Lett. \textbf{74}, 3451 (1995).

\bibitem{Tanaka00} S.~Kashiwaya and Y.~Tanaka, Rep. Prog. Phys. \textbf{63}, 1641 (2000).

\bibitem{tanaka10} Y.~Tanaka, Y.~Mizuno, T.~Yokoyama, K.~Yada, and M.~Sato, Phys. Rev. Lett. \textbf{105}, 097002 (2010).

\bibitem{takami} S.~Takami, K.~Yada, A.~Yamaka, M.~Sato, and Y. Tanaka, J. Phys. Soc. Jpn. \textbf{83}, 064705 (2014).

\bibitem{braunecker10} B.~Braunecker, G.~I.~Japaridze, J.~Klinovaja, and
  D.~Loss, Phys. Rev. B. \textbf{82}, 045127 (2010).

\bibitem{klinovaja12a} J.~Klinovaja and D.~Loss, Phys. Rev. B
  \textbf{86}, 085408 (2012).

\bibitem{klinovaja12b} J.~Klinovaja, P.~Stano, and D.~Loss,
  Phys. Rev. Lett. \textbf{109}, 236801 (2012).
	
\bibitem{eugene} E. Dumitrescu, B. Roberts, S. Tewari, Jay D. Sau, and S. Das Sarma, Phys. Rev. B \textbf{91}, 094505 (2015).

\bibitem{takei} S. Takei, B. M. Fregoso, H.-Y. Hui, A. M. Lobos, and S. Das Sarma, Phys. Rev. Lett. \textbf{110}, 186803 (2013).

\bibitem{cheng09} M. Cheng, R. M. Lutchyn, V. Galitski, and S. Das Sarma, Phys. Rev. Lett. \textbf{103}, 107001 (2009).

\bibitem{cheng10} M. Cheng, R. M. Lutchyn, V. Galitski, and S. Das Sarma, Phys. Rev. B. \textbf{82}, 094504 (2010).

\bibitem{dassarma12} S. Das Sarma, Jay D. Sau and T. D. Stanescu, Phys. Rev. B \textbf{86} 220506(R) (2012).

\end{thebibliography}
\end{document}